\documentstyle[aps,pre,multicol,psfig]{revtex}
\begin{document}
\draft 

\title{Equilibrium Shapes and Faceting for Ionic Crystals of 
Body-Centered-Cubic Type}
\author{Enrico Carlon}
\address{INFM - Dipartimento di Fisica, Universit\'a di Padova,
I-35131 Padova, Italy}
\author{Henk van Beijeren}
\address{Institute for Theoretical Physics, Utrecht University,
P.O. Box 80006, 3508 TA Utrecht, The Netherlands}

\date{\today}

\maketitle

\begin{abstract}
A mean field theory is developed for the calculation of the surface free 
energy of the staggered BCSOS, (or six vertex) model as function of the 
surface orientation and of temperature. The model approximately describes 
surfaces of crystals with nearest neighbor attractions and next nearest 
neighbor repulsions. The mean field free energy is calculated by expressing 
the model in terms of interacting directed walks on a lattice. 
The resulting equilibrium shape is very rich with facet boundaries and 
boundaries between reconstructed and unreconstructed regions which can be 
either sharp (first order) or smooth (continuous). In addition there are 
tricritical points where  a smooth boundary changes into a sharp one and 
triple points where three sharp boundaries meet.
Finally our numerical results strongly suggest the existence of conical 
points, at which tangent planes of a finite range of orientations all 
intersect each other.
The thermal evolution of the equilibrium shape in this model shows strong 
similarity to that seen experimentally for ionic crystals.
\end{abstract}

\begin{multicols}{2} \narrowtext

Crystals in thermal equilibrium are typically composed of flat 
regions (facets) corresponding to high symmetry directions in the 
crystal lattice and possibly some rounded parts. 
At sufficiently low temperatures the equilibrium crystal shape is 
dominated by the facets, while with increasing temperature more 
and more rounded regions occupy larger areas of the crystal surface.
At a given temperature a facet may shrink completely and disappear 
from the equilibrium shape; this corresponds to a {\em roughening} 
transition, which is characterized microscopically by the vanishing of 
the step free energy on the facet \cite{henkim}.

Another interesting phenomenon occurring on crystal surfaces is that 
of {\em faceting}, or phase separation of unstable orientations 
\cite{henkim,williams}. An orientation is unstable (or metastable) when 
the total surface free energy of the crystal can be lowered by replacing 
that orientation by a combination of other orientations, connected to each 
other under sharp edges, with an average orientation equal to the original 
one.
This process is similar, to that of liquid-gas phase separation,
where the free energy of the system at a given homogeneous density
can be lowered by combining the free energies of a denser liquid
phase and a more dilute gas phase \cite{notea}.
In the case of crystal surfaces the step density is the equivalent
of the particle density in the liquid-gas phase separation.

A third phenomenon observed frequently on crystal surfaces is {\em 
reconstruction}, implying that the unit cell of the surface in equilibrium
is larger than that obtained by making a section through the bulk crystal
structure. The most common cause for this is lowering of surface
energy due to the rearrangement of atoms in the surface layer.
Raising temperature often destroys reconstructions
at a {\em deconstruction temperature} as a result of increasing entropy.
In many cases there is a subtle interplay between reconstruction and
the other two phenomena.

In this paper we describe the thermal evolution of the equilibrium
shape for a model describing equilibrium surfaces of crystals of
body-centered-cubic (bcc) type. Using a mean-field approximation we 
calculate the surface free energy as function of the surface orientation 
and temperature.
The model has a rich phase diagram showing roughening, faceting and
deconstruction transitions and it provides an approximate description of
ionic crystals of CsCl type, where ions of the same type repel each other,
while there is a strong attraction between nearest neighbors,
which are oppositely charged. 

The paper is organized as follows: In Section\ \ref{sec:model}, we 
introduce the model, in Section \ \ref{sec:mf} we present the mean-field 
approximations that we employ to calculate the surface tension of the 
crystal as function of surface orientation and temperature.
In Section\ \ref{sec:ecs} we discuss the evolution of the equilibrium crystal 
shape as function of the temperature.
In Section\ \ref{sec:disc} we summarize the results obtained and make a
comparison with known models and experiments showing similar features.

A preliminary account of this work has been presented already in 
Ref. \cite{rapid}. Here we develope a different mean-field theory
which has several advantages over that of the previous approach.

\section{The model}
\label{sec:model}

We consider a bcc crystal composed of two different types of atoms, 
say A and B, which occupy the sites of the two interpenetrating cubic
lattices that form the bcc structure.
In the solid-on-solid (SOS) approximation the surface configurations
are given by integers $h_i^A$ and $h_j^B$ describing the heights of
the surface atoms with respect to a reference plane. $h_i^A$ and 
$h_j^B$ are odd and even integers, respectively.

We consider the following Hamiltonian \cite{knops}:

\begin{eqnarray}
H = J_0 \sum_{\langle ij \rangle} \left( \left|h_{i}^{A} - 
h_{j}^{B} \right| - 1 \right) + 
\nonumber \\
\frac{\epsilon}{2}\sum_{(kl)} 
\left| h_{k}^{A} - h_{l}^{A}\right| + 
\frac{\epsilon + 2 \delta }{2}\sum_{(mn)} 
\left| h_{m}^{B} - h_{n}^{B}\right|
\label{hamilt1}
\end{eqnarray}
where the sums are constrained to neighboring AB, AA and BB couples 
($\langle . \rangle$ and $(.)$ denote summations over the nearest 
respectively the next nearest neighbors).

\begin{figure}[b]
\centerline{
\psfig{file=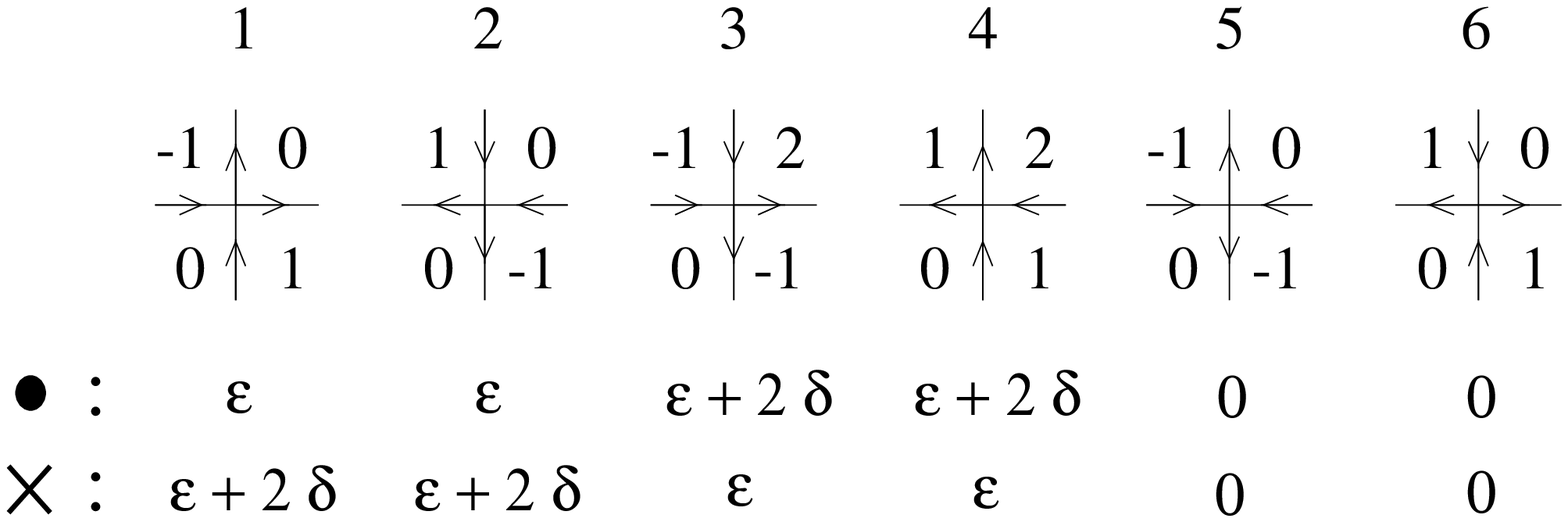,height=2.8cm}}
\vskip 0.2truecm
\caption{The six vertices and their energies in the two
distinct sublattices ($\bullet$ and $\times$). At each vertex, 
the four integers denote possible height variables.}
\label{FIG01}
\end{figure}

We take the limit $J_0 \to \infty$ so that height differences between 
neighboring atoms are restricted to their minimal value ($\pm 1$) and the 
model can be mapped onto a six vertex model \cite{henkPRL}.
The correspondence between vertices and height variables is shown in
Fig.\ \ref{FIG01}. Each vertex satisfies the ice rule, requiring
that two of its arrows point inward and two point outward. We stress 
that if $\delta \neq 0$ the model is mapped onto a {\em staggered} six 
vertex model with the vertex lattice subdivided into two sublattices 
on which the vertices 1,\ldots,4 have different 
energies as shown in Fig.\ \ref{FIG01}.
As the two representations are completely equivalent in the rest of the 
paper we will sometimes use the term BCSOS (body centered solid-on-solid) 
model and sometimes six vertex model.

We take $\epsilon < 0$ and $0 < \delta \ll -\epsilon$. This model may 
give a good approximate description of ionic crystals where A and B are 
ions with opposite charges. With our choice of energies neighboring A and 
B atoms strongly attract each other ($J_0 \to \infty$ in the model) and 
atoms of the same type repel each other ($\epsilon$,$\epsilon + 2 \delta < 0$) 
and so the model may be expected to give at least a good qualitative picture
of real ionic crystals, even though further neighbor interactions are
ignored. By choosing $\delta > 0$ we assumed that on top of the Coulomb
repulsion between equal species there is some other contribution to the 
Hamiltonian, which makes the interaction energies between next nearest 
neighbor AA and BB couples slightly different. 

\begin{figure}[b]
\centerline{
\psfig{file=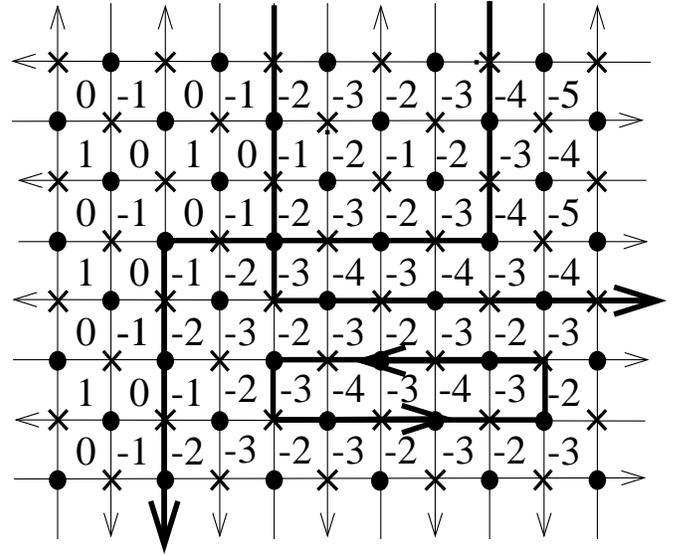,height=7.5cm}}
\vskip 0.2truecm
\caption{Connected walks of reversed arrows (thick lines) denote excitations 
with respect to the Manhattan background (thin lines). Integers are height 
variables in the BCSOS representation.}
\label{FIG02}
\end{figure}

The ground state, when $\epsilon$ is negative, describes a $c(2\times 2)$ 
reconstructed $(001)$ facet with one component (B) at constant height
(say, $h_j^B = 0$), and the other (A) with alternating heights 
($h_i^A = \pm 1$) above and below the first one.
In the vertex lattice all the arrows point alternatingly up and
down or left and right, a configuration which we will refer to
as ``Manhattan lattice".
Figure \ref{FIG02} shows some elementary excitations of the ground state
(thick lines denote excitations with respect to the Manhattan ground 
state configuration indicated by thin lines). Reversed arrows on a closed 
loop produce a closed terrace of surface atoms two lattice units higher 
(or lower) than their ground state heights. A connected path of reversed 
arrows running between two boundaries of the Manhattan lattice, corresponds 
to a step on the $(001)$ facet for the BCSOS model. 
On the six vertex lattice such a path can be described as a self avoiding 
walk that cannot visit the same bond twice, but is allowed to cross itself.
In reversing arrows the step/walk has to preserve the ice-rule and this 
limits its possible trajectories. At each non-crossing site, the walk has 
two options: either to go straight, maintaining its previous direction,
with an energy cost of $2 \delta$, or to turn towards the direction allowed 
by the Manhattan lattice, which costs an energy $-\epsilon$ (since we 
consider the case $0 < \delta \ll - \epsilon$, typically walks will be 
composed of very long segments with rare turns).
Steps can cross each other, as shown in Fig.\ \ref{FIG02}; step
crossings are energetically favored: at a crossing point a ground
state vertex is replaced by a vertex with the same energy with
all four arrows reversed. There is a gain in energy of $4 \delta$,
in comparison to the energy the system would have to pay for two straight
non-crossing segments of unit length.

The six vertex model describes not only the $(001)$ surface of a bcc 
crystal, but also all the side orientations $(ts1)$ with $| t | + | s | 
\leq 1$.  Given a configuration of vertices, the horizontal and vertical
polarizations $q$ and $p$ are defined by
\begin{eqnarray}
q = n_{\uparrow} - n_{\downarrow}  \,\,\,\,\,\,\,\,\,\,\,\,\,
p = n_{\rightarrow} - n_{\leftarrow}
\label{polarizations}
\end{eqnarray}
where $n_{\uparrow}$, $n_{\downarrow}$, $n_{\rightarrow}$,
$n_{\leftarrow}$ are the densities of up, down, left and right
arrows respectively.
A {\it fully polarized} state, namely a state with all arrows pointing
(say) up and to the right describes a $(011)$ facet of the bcc
crystal. 
The relationships between the variables $t,s$ and $p,q$ are given by 
$t=(p+q)/2$, $s=(q-p)/2$. Please note we have chosen the principal axes of
the crystal under angles of $45^\circ$ with the principal axes of the vertex 
lattice.

\begin{figure}[b]
\centerline{
\psfig{file=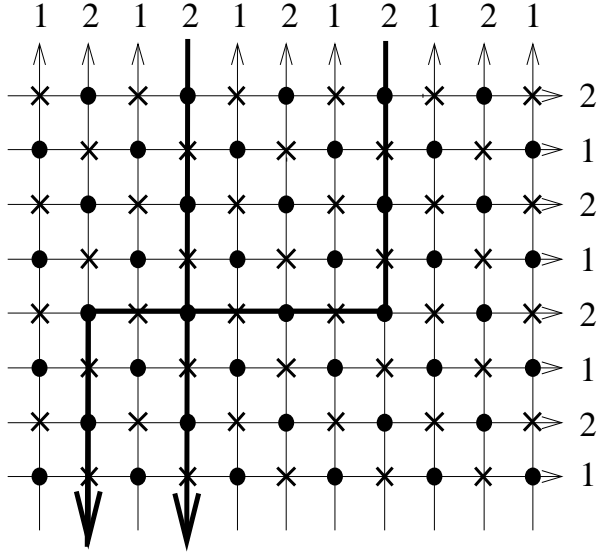,height=7.5cm}}
\vskip 0.2truecm
\caption{Excitations (thick lines) with respect to the fully polarized 
state $p=q=1$, corresponds to steps on the $(011)$ facet. We distinguish 
between even and odd columns and rows.}
\label{FIG03}
\end{figure}

Figure \ref{FIG03} shows excitations on $(011)$ type facets. In this case
no closed loops are possible, since they would violate the ice-rule,
and excitations can be described as {\em directed walks} on the lattice, 
i.e. walks that can only step down or to the left.
Two walks cannot visit the same lattice bond, but they can cross each 
other, as shown in the figure. The mean-field calculation of the free 
energy of the six vertex model presented in this paper takes as starting 
point the fully polarized state $q=p=1$. The rows and columns are subdivided 
into even and odd ones as in Fig. \ref{FIG03}; the polarizations $p$ and 
$q$ are given by
\begin{eqnarray}
p = \frac{p_1 + p_2}{2}
\,\,\,\,\,\,\,\,\,\,
{\rm and}
\,\,\,\,\,\,\,\,\,\,
q = \frac{q_1 + q_2}{2}
\end{eqnarray}
in terms of the sublattice polarizations $p_1$, $p_2$, $q_1$ and $q_2$.
In states with $q = q_1 = q_2$ and $p = p_1 = p_2$ arrows are reversed 
with respect to the fully polarized state with equal probabilities on the 
two sublattices, i.e. the surface is unreconstructed.
If on the other hand one finds states with $q_1 \neq q_2$ or $p_1 \neq p_2$, 
the surface is reconstructed. Since the energy of a configuration is 
invariant under the interchange of sublattices 1 and 2 (or equivalently a 
translation over the lattice vector (1,1) on the vertex lattice), each 
reconstructed state is degenerate with another one with the values of 
$p_1,q_1$ and $p_2,q_2$ interchanged.
For either of the states the 1-2 exchange symmetry is spontaneously broken.
For example, the two Manhattan states are given by $q_1 = p_1 = -q_2 = 
-p_2 =\pm 1$. 

\section{Calculation of the mean-field free energy}
\label{sec:mf}

\subsection{Energy and ground state properties}
\label{subsec:gsp}

It is instructive, before presenting the details of the mean-field approach, 
to consider the ground state properties of the model, as this will provide 
already important information on the low temperature behavior of the system.
Throughout the rest of the paper we set the ground state energy of the 
Manhattan state equal to zero by a shift of all the vertex energies over 
an amount $\epsilon$, so the lowest vertex energy becomes zero.
With this convention the energy per site of the fully polarized state 
(where all arrows point, say, up and to the right) becomes $\delta$. 
We will take this state as starting point for our mean field calculations. 
All other allowed vertex configurations can be represented by a set of 
directed walks on the lattice, as illustrated in Fig. \ref{FIG03}. The 
lattice points will be distinguished into four different types $ij$, with 
$i,j=1,2$. 
The 11 points for example will be the crossing points of odd rows with odd
columns of arrows. For a given configuration of directed walks $c_{ij}$ 
and $t_{ij}$  will indicate the total numbers of crossings respectively turns 
at sites of type $ij$.

For instance, for the configuration of Fig. \ref{FIG03} one has $c_{22} 
= 1$, $c_{11} = c_{12} = c_{21} = 0$ and $t_{22} = 2$, $t_{11} = t_{12} 
= t_{21} = 0$.
An isolated straight path of reversed arrows on the fully polarized state
does not change the energy, since it is formed by a collection of vertices 
with alternating energies $2 \delta$ and $-2 \delta$, which sum up to zero.
However, crossings and turns do contribute to the energy. A given configuration 
with $c_{ij}$ crossings and $t_{ij}$ turns at sites of type $ij$ has total 
energy:
\begin{eqnarray}
E \left( c_{ij}, t_{ij} \right) = N^2 \delta - 4 \delta \left( c_{11} + 
c_{22} \right) + 4 \delta \left( c_{12} + c_{21} \right) \nonumber \\
- \epsilon \left( t_{11} + t_{22} \right)
- \left(\epsilon + 2 \delta \right) 
\left( t_{12} + t_{21} \right),
\label{energy}
\end{eqnarray}
since for crossings on $11$ or $22$ sites one has a gain of energy $4 \delta$,
crossings on $12$ or $21$ sites cost an energy $4 \delta$, while turns on 
sites of types $11$, $22$ or $12$, $21$ cost energy $-\epsilon$ and
$-\epsilon-2 \delta$ respectively.
 
This expression indeed becomes minimal for the Manhattan states, where one 
has $c_{11}=c_{22}=N^2/4$ and $c_{12}=c_{21}=t_{ij}=0$. One may look for 
the minimal energy of states with a homogeneous distribution of polarization 
on each of the sublattices. For given sublattice polarizations $q_i,p_i$ 
the number of crossings may be estimated as
\begin{eqnarray}
c_{ij} = \frac{N^2}{16} \left(1- p_i \right) \left(1- q_j \right),
\label{crossings}
\end{eqnarray}
an expression that becomes {\it exact} if no turns are allowed at all (as it 
happens at zero temperature). Inserting this value into Eq. (\ref{energy}) 
and minimizing with respect to $q_i$ and $p_i$ at fixed $0 \leq q,p \leq 1$ 
one finds the homogeneous ground state energy per vertex:
\begin{eqnarray}
\epsilon_0(p,q) &=& \delta \left(  p  +  q  \right) - \delta  p q 
\label{groundst}
\end{eqnarray}
which is obtained either for $q_1 = 2 q -1$, $p_1 = 2 p - 1$, $p_2 =
q_2 = 1$ or for $q_1 = q_2 = 1$, $p_2 = 2 p - 1$, $q_2 = 2 q - 1$.
These solutions describe doubly degenerate reconstructed surfaces with
the Manhattan states obtained in the limit $q,p \to 0$.

\begin{figure}[h]
\centerline{
\psfig{file=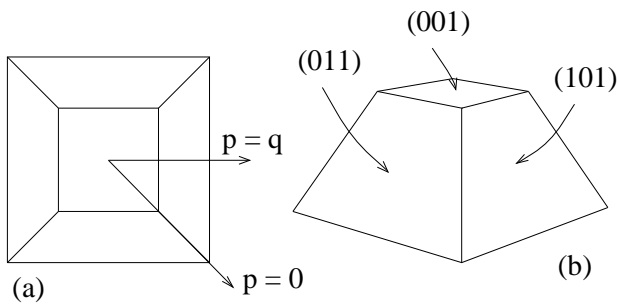,height=4cm}}
\vskip 0.2truecm
\caption{The equilibrium crystal shape at $T=0$.}
\label{FIG04}
\end{figure}

However, the homogeneous ground state energy (Eq.  (\ref{groundst}))
is a non-convex function of $p$ and $q$.
It is well known (see, for instance, \cite{henkim}) that instabilities
arise when the free energy per unit of projected area is a non-convex
function of the surface orientation. These will give rise to faceting of 
surfaces with orientations in a non-convex range. The equilibrium shape
constructed from (\ref{groundst}) is shown in Fig.\ \ref{FIG04};
it consists of flat facets only: the ``top" $(001)$ and the
``side" facets $(011)$, $(101)$, $(0\bar{1}1)$ and $(\bar{1}01)$.

Notice that Eq. (\ref{groundst}) implies that instabilities ought 
to persist over some range of temperatures, where entropic effects are 
not sufficiently strong to turn the concave free energy into a convex 
(i.e. stable) one. The only exception to this will be found at the
lines $q=0$ or $p=0$, where the bilinear term in Eq.~(\ref{energy}) vanishes.
These lines correspond to orientations between the $(101)$ and $(011)$ 
facets of Fig.\ \ref{FIG04}. It is natural then to expect that finite 
temperature effects (i.e. the appearance of rounded regions) will first 
manifest themselves along the common edges of these facets.
Summarizing, these simple considerations related to the $T=0$
properties of the system lead us to conclude that: \begin{itemize}
\item[(1)] Sharp edges between the $(001)$ and $(011)$ facets will 
persist at finite temperatures
\item[(2)] The corners between the facets and the edges between 
the $(011)$ type facets will probably become rounded first.
\end{itemize}

\subsection{Entropy and free energy}
\label{subsec:mf}
In the sequel we will focus on the behavior of the system for $\delta \ll - 
\epsilon$ and restrict ourselves to not too high temperatures, when directed 
walks are typically composed of long straight segments, because the Boltzmann 
weight for a turn $\exp (\beta\epsilon)$, is very small. 
As a result of this, different walks can hardly develop any local 
correlations, and the mean-field analysis should be very accurate.

For given sublattice polarizations $p_i,q_i$ the numbers of crossings 
$c_{ij}$ in a homogeneous state can still be estimated to be given by 
Eq.~(\ref{crossings}), thanks to the absence of correlations between 
directed walks. The energy of such a state is then given by 
Eq.~(\ref{energy}), with the $c_{ij}$ following from Eq.~(\ref{crossings}). 
To obtain the free energy at nonzero temperature we need both the average 
numbers of turns $t_{ij}$ and the entropy of the system.
In our mean field approximation both are obtained from the same calculation.

Let us consider as an example the contribution to the entropy resulting 
from the $t_{11}/2$ turns from a column $1$ to a row $1$. There are $N^2 
(1 - q_1)/8$ inverted arrows on type-1 columns pointing towards a 11-site
and $c_{11}$ of those are occupied by crossings. There are thus:
\begin{eqnarray}
\Gamma_{11}^{\rm c \to \rm r} = \frac{N^2}{8} (1- q_1) - c_{11}.
\end{eqnarray}
sites available for the $t_{11}/2$ turns. 
The superscript $\rm c \to \rm r$ indicates that we are considering 
turns from columns to rows. For turns from rows to columns one finds 
analogously:
\begin{eqnarray}
\Gamma_{11}^{\rm r \to \rm c} = \frac{N^2}{8} (1- p_1) - c_{11}.
\end{eqnarray}
Both of these equations are readily generalized for turns from rows/columns
to columns/rows of either type. Neglecting again correlations, one finds 
that the number of possible ways of making the turns from a column to a row 
on sites $ij$ is then given by the binomial coefficient of $\Gamma^{c \to 
r}_{ij}$ and $t_{ij}/2$.
Collecting the contributions from all possible turns one finds:
\begin{equation}
S=k_B \sum_{ij} \left[ \ln \left(
\begin{array}{c}
\Gamma_{ij}^{\rm c \to \rm r} \\
t_{ij}/2
\end{array}
 \right)+
\ln
 \left(
\begin{array}{c}
\Gamma_{ij}^{\rm r \to \rm c} \\
t_{ij}/2
\end{array}
 \right)\right],
\label{entropy}
\end{equation}
where $k_B$ is Boltzmann's constant.

Combining Eqs. (\ref{energy}), (\ref{crossings}) and (\ref{entropy}) one finds 
the total free energy $F = E - T S$ as function of the parameters $p_i$, $q_i$ 
and $t_{ij}$. It is convenient to minimize first with respect to the 
numbers of turns. 
The equations $\partial F/\partial t_{ij} = 0$ for $i,j = 1,2$ yield to 
lowest order in $e^{\beta \epsilon}$:
\begin{equation}
t_{ij} = \frac{N^2}{4} \,\, \sqrt{(1-p_i^2)(1-q_j^2) } \,\,
e^{\beta(\epsilon +2\delta \delta_{ij})},
\label{n_ij}
\end{equation}
with $\delta_{ij}$ a Kronecker delta.
The dependence on the number of turns $t_{ij}$ can thus be eliminated, and 
to lowest order in $e^{\beta \epsilon}$ the free energy per site becomes:
\begin{eqnarray}
f(q_1,q_2,p_1,p_2) = \delta \left[1-\frac 1 4 (q_1-q_2)(p_1-p_2) \right] - 
\nonumber \\
\frac{1}{8 \beta} \sum_{ij} e^{\beta (\epsilon+2\delta \delta_{ij})}
\sqrt{(1-p_i^2)(1-q_j^2)}
\label{fretomin}
\end{eqnarray}
It is more convenient to express the sublattice polarizations as:
\begin{eqnarray}
\left\{
\begin{array}{ccc}
q_1 & = & q - \alpha \\
q_2 & = & q + \alpha 
\end{array}
\,\,\,\,\,\,\,\,\,\,\,
\,\,\,\,\,\,
\right\{
\begin{array}{ccccc}
p_1 & = & p - \gamma \\
p_2 & = & p + \gamma
\end{array}
\end{eqnarray}
where $p$ and $q$ are the average slopes of the surface and $\alpha$ 
and $\gamma$ can be interpreted as order parameters for the reconstructed 
state; if they are both zero the polarizations $q_1$, $p_1$ and $q_2$, 
$p_2$ of the two sublattices are identical and the surface is in an
unreconstructed state.

The mean field free energy can be found by 
minimization of (\ref{fretomin}) for fixed $q$ and $p$: 
\begin{eqnarray}
f_{\rm MF}(p,q) = \min_{\left\{\alpha, \gamma \right\} }
f(q - \alpha, q + \alpha, p - \gamma, p + \gamma)
\label{mfree}
\end{eqnarray}
The unreconstructed state corresponding to $\alpha = \gamma = 0$ has a 
free energy given by:
\begin{equation}
f^{(u)}_{\rm MF}(q,p) = \delta - C(\beta)\sqrt{(1-p^2)(1-q^2)},
\label{freeunr}
\end{equation}
with
\begin{equation}
C(\beta)= \frac{1}{4 \beta} e^{\beta \epsilon} 
\left(e^{2 \beta \delta} + 1\right).
\label{C}
\end{equation}
The solution with $\alpha = \gamma = 0$ is always a stationary point of the 
free energy of Eq. (\ref{fretomin}), however it is not always a {\it global} 
free energy minimum as we will see. For arbitrary $p$ and $q$ we implemented 
the minimization numerically, except along some symmetry lines where it is
possible to solve the problem analytically. We will start discussing 
these special cases first.

\subsubsection{The orientations $0 \leq p = q \leq 1$}

For $q=p$ we take $\alpha = \gamma$ in Eq. (\ref{mfree}) and the free energy 
to be minimized takes the form:
\begin{eqnarray}
f(\alpha) = \delta (1 - \alpha^2) - \frac{1}{4 \beta} e^{\beta \epsilon}
\left\{  e^{2 \beta \delta} \left[1 - (q^2 + \alpha^2 )\right]
+ \right. \nonumber\\
\left. 
\sqrt{ \left[ 1 - (q - \alpha)^2 \right] \left[ 1 - (q + \alpha)^2 \right] }
\right\}
\label{freea}
\end{eqnarray}
with $q - 1 \leq \alpha \leq q + 1$.

Setting $\partial f/\partial \alpha = 0$ one gets the following equation:
\begin{eqnarray}
& 2 \alpha & \left\{
\delta - \frac{e^{\beta (\epsilon + 2 \delta)}}{4 \beta}
- \frac{e^{\beta \epsilon}}{4 \beta} \frac{1 - (\alpha^2 - q^2)}
{\sqrt{\left[ 1 - (\alpha - q)^2 \right] \left[ 1 - (\alpha + q)^2 
\right]}} \right\}
\nonumber \\
&=& 0
\label{zeroeq}
\end{eqnarray}
The solution $\alpha = 0$ of course corresponds to the unreconstructed state, 
for which on the symmetry line $q=p$ the free energy takes a simple parabolic 
shape:
\begin{eqnarray}
f^{(u)}_{\rm MF}(q,q) = \delta - C(\beta) + C(\beta) \, q^2
\label{unrecq=p}
\end{eqnarray}
The other possible solution of Eq. (\ref{zeroeq}) can be found by defining
first $\Delta \equiv \alpha^2 - q^2$; squaring Eq. (\ref{zeroeq}) one obtains 
a quadratic equation for $\Delta$ with solution:
\begin{eqnarray}
\Delta = 1 - q \frac{2 \delta - C(\beta) - \tilde{C} (\beta) }
{\sqrt{[\delta - C(\beta)] [\delta - \tilde{C}(\beta)]}}
\label{delq=p}
\end{eqnarray}
with $\tilde{C} (\beta) = e^{\beta \epsilon} (e^{2 \beta \delta} 
- 1)/4 \beta$.
This equation can also be rewritten in the following form:
\begin{eqnarray}
\alpha^2 = \left(q_0 - q \right) \left(\frac{1}{q_0} - q\right)
\label{alq=p}
\end{eqnarray}
with:
\begin{eqnarray}
q_0 = \sqrt{ \frac{\delta - C(\beta)}
{\delta - \tilde{C}(\beta)} } \leq 1
\label{deftildeq0}
\end{eqnarray}
Notice that the reconstructed solution with $\alpha \neq 0$ exists only
for a limited range of temperatures and orientations, namely for $\delta
> C(\beta)$ and $|q| \leq {q}_0$,
where the right hand side of Eq. (\ref{alq=p}) and the argument of
the square root of Eq. (\ref{deftildeq0}) are both positive.
Notice also that as $q \to q_0^-$ the reconstruction order parameter
$\alpha$ vanishes as $\alpha \sim \sqrt{q_0 - q}$, i.e. with the mean-field
exponent $1/2$.

Substituting the value of $\Delta$ given by (\ref{delq=p}) into 
Eq. (\ref{freea}) one finds for the reconstructed free energy:
\begin{eqnarray}
f^{(r)}_{\rm MF}(q,q) &=&
2 \sqrt{ [\delta-C(\beta)][\delta- \tilde{C}(\beta)]} \,\, q  
\nonumber\\ &-& 
[\delta - C(\beta) - \tilde{C} (\beta) ] 
\,\, q^2
\label{recq=p}
\end{eqnarray}
In the range of parameters where the reconstructed solution exists this 
has always lower free energy than the unreconstructed solution of 
Eq. (\ref{unrecq=p}).
At zero temperatures ($\beta \to + \infty$) the reconstructed free energy 
(\ref{recq=p}) yields the exact ground state energy obtained in 
Eq.(\ref{groundst}). For $q=q_0$ the reconstructed (\ref{recq=p}) and 
unreconstructed (\ref{unrecq=p}) free energies take the same value with
equal derivatives.

\begin{figure}[ht]
\centerline{
\psfig{file=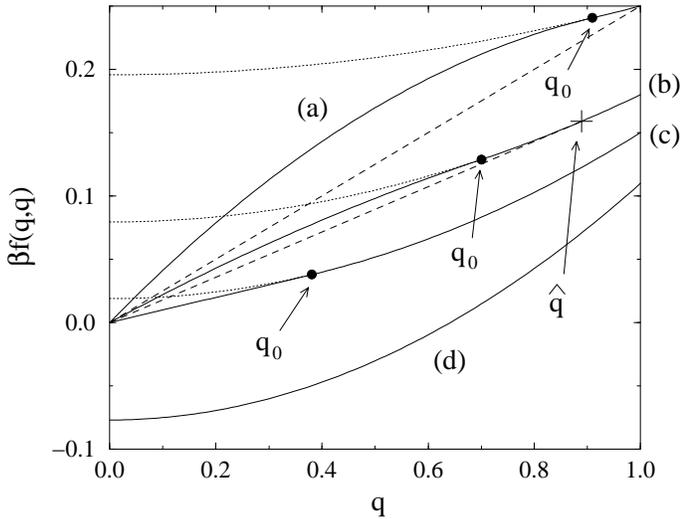,height=7cm}}
\vskip 0.2truecm
\caption{
(Solid lines) Mean field free energies as function of the slope $q$ 
calculated along the direction $p=q$ and for four different temperatures 
with $\epsilon = -2.0$, $\delta = 0.2$: (a) $T = 0.80$, (b) $T = 1.111$, 
(c) $T =1.333$ and (d) $T = 1.818$. (Dotted lines) Analytic continuation 
of the unreconstructed free energy in the region $q < q_0$. 
(Dashed lines) Maxwell construction connecting stable orientations.
}
\label{FIG05}
\end{figure}

Figure \ref{FIG05} shows the free energies (solid lines) along the line 
$p=q$ for different values of the temperature. The dotted lines are the 
analytical continuations of the unreconstructed free energy in the regions 
($q < q_0$) where this is not the absolute minimum of the free energy 
(\ref{fretomin}). For sufficiently large temperatures (e.g. curve (d) in 
Fig. \ref{FIG05}) the absolute minimum always corresponds to an 
unreconstructed surface.
Notice that at low temperatures the reconstructed free energy (\ref{recq=p}) 
is concave as we expected from the analysis of the ground state energy 
(\ref{groundst}). 

Stable orientations can be found from the double tangent or Maxwell
construction which is shown as a dashed line in the figure. In case (a) the
only stable orientations along the $q=p$ line are $q=0$ and $q=1$. In case (b) 
the Maxwell construction connects the orientations $q = 0$ with $q = {\hat q}$,
where ${\hat q}$ is given in Eq. (\ref{defq1}). Therefore only the orientations
in the range ${\hat q} \leq q \leq 1$ and $q = 0$ are stable.

Above a temperature $T_s$ obtained from the condition:
\begin{eqnarray}
\delta = C(\beta_s) + \tilde{C}(\beta_s)
\label{tsmooth}
\end{eqnarray}
the reconstructed free energy (\ref{recq=p}) becomes stable, (see curve (c) 
in the figure). In this case all the surface orientations along the line 
$q=p$ are stable and in the range $0 \leq q \leq q_0$ these are reconstructed.

The properties of this free energy and of the equilibrium shape along the 
symmetry axis $p=q$ will be discussed in more detail in the next section.

\subsubsection{The orientations $p = 0$, $0 \leq q \leq 1$}

For $p=0$ we minimize the free energy (\ref{mfree}) with respect to the 
parameters $\alpha$ and $\gamma$. The solution takes the form:
\begin{eqnarray}
\alpha^2 &=& \left(\tilde{q}_0 - q \right) \left(\frac{1}
{\tilde{q}_0} - q\right) \label{alp=0}\\
\gamma^2 &=& \frac{1}{\tilde{q}_0} \frac{\tilde{q}_0 - q}{1 - q 
\tilde{q}_0}
\end{eqnarray}
with:
\begin{eqnarray}
\tilde{q}_0 &=& \frac{1}{\delta} \sqrt{\delta^2 - C^2(\beta)}.
\end{eqnarray}
This yields the following value for the reconstructed free energy:
\begin{eqnarray}
f^{(r)}_{\rm MF}(0,q) = \sqrt{\delta^2 - C^2(\beta)} \, q
\label{recp=0}
\end{eqnarray}
which is valid for $q \leq \tilde{q}_0$ and $\delta > C(\beta)$.

This free energy is linear as function of the slope parameter $q$.
The numerical analysis of the free energy for small values of $q$
shows that the reconstructed orientations with small $p$ are not stable,
implying that the equilibrium crystal shape has sharp edges along this
orientation as we will discuss in detail in Sec. \ref{sec:ecs}.

\subsubsection{The orientations $q = 1$, $0 \leq p \leq 1$}

The exact value of the free energy at $q=1$ can be calculated easily. 
We show here that in this limiting case the mean-field free energy
reproduces the exact result. Let us consider a vertex lattice of size 
$N \times N$: $q=1$ describes a set of orientations where all the 
vertical arrows are up, in a state of maximal polarization (the 
state is composed of vertices 1 and 4).
The horizontal arrows can point either to the left or to the right,
but once a horizontal arrow at a boundary is fixed, all the arrows 
connected to it along a horizontal line point into the same direction, 
otherwise the ice rule would be violated. The average energy
per site is $\delta$ since along horizontal lines, there is an
alternating sequence of vertices of energies $0$ and $2 \delta$.
The total entropy of the $N \times N$ lattice is of order $N$ since
the vertical arrows are frozen and the horizontal ones along a row are 
identical. Therefore the free energy {\em per site} becomes:
\begin{eqnarray}
f(p,q=1) &=& \delta + O(1/N)
\end{eqnarray}
and in the thermodynamic limit, $N \to \infty$, the entropic term
can be neglected. This limiting value of the free energy is reproduced
correctly by our mean-field calculations at all temperatures.
The free energy for vicinal orientations very close to the $(011)$
type facets can also be calculated exactly. It just involves calculating 
the free energy of isolated steps on these facets. This calculation is 
worked out in the appendix. It turns out that our mean field approximation 
also describes the free energy of these vicinal orientations correctly. 
This is no surprise, since for these orientations crossings are very rare and 
unimportant, so that our method for estimating the entropy and the numbers of 
turns becomes exact.

\subsubsection{The orientation $p=q=0$}

The other limiting case to be considered is the orientation with $q = p =0$. 
In this case for $\delta > C(\beta)$ (\ref{mfree}) yields two minima with 
$\alpha = \gamma = \pm 1$, which correspond to the two {\em reconstructed} 
Manhattan states with $f^{(r)}_{\rm MF}(p=0,q=0) = 0$. Since the ground state 
energy of the system equals zero we expect that in reality $f(p=0,q=0)<0$ at 
finite temperatures. The reason we do not find this in our mean field 
approximation is due to the fact that we have neglected correlations between 
directed walks. E.g.\ at very low temperatures the most important excitations 
of the ground state are closed loops around an elementary square on the 
Manhattan lattice. These obviously exhibit strong correlations; four turns 
are located on the corners of the same elementary square. However, their 
contribution to $f$ is of the order $\exp(4\beta \epsilon)$ and therefore 
ignoring these terms is consistent within our approximation scheme.
Above the temperature $T_{\rm MF}=1/({k_B \beta_{\rm MF}})$ satisfying
\begin{equation}
\delta = C(\beta_{\rm MF})
\label{Tmf}
\end{equation}
the minimum of (\ref{fretomin}) corresponds to $\alpha = \gamma = 0$
and the orientation $p=q=0$ is unreconstructed, with free energy given
by (\ref{freeunr}), which is negative as expected.
So for this orientation our mean field theory predicts a phase transition
where the reconstruction order parameters ($\alpha$ and $\gamma$) jump from 
$\pm 1$ to zero and the step free energy vanishes with a mean-field exponent 
$1/2$ as can be seen from the free energies (\ref{recq=p}) and (\ref{recp=0}).

Transfer matrix calculations \cite{PRB} strongly suggest that in reality the 
reconstruction transition and the roughening transition occur at separate
temperatures and that both are continuous. However, the temperature difference 
between the two transitions decreases very rapidly with increasing $|\epsilon|$
(our estimates \cite{PRB} yield an approach as $e^{12\beta\epsilon}$) and, as 
$|\epsilon|$ gets larger the increase of $\beta f$ from practically zero to 
appreciable values occurs within an increasingly narrow range of 
$\beta \epsilon$. 
Since we expect our mean field approximation to become exact in the limit 
$|\epsilon|/\delta \to \infty$, the predicted coincidence of roughening and 
deconstruction as well as the first order character of this phase transition 
are in fact to be expected.

\subsubsection{Properties of the free energy for arbitrary $p$ and $q$}

At the points $p=q=q_0$ and $p=0$, $q=\tilde{q}_0$ the reconstruction 
parameter $\alpha$ becomes zero, as follows from Eqs. (\ref{alq=p}) 
and (\ref{alp=0}). One can find the locus $\alpha=0$ for general $p$ and $q$
by setting $q_1 = q - \alpha$, $q_2 = q + \alpha$ and $p_1 = p - \gamma$, 
$p_2 = p + \gamma$ and requiring that the matrix of the second derivatives
of the free energy with respect to the parameters $\alpha$ and $\gamma$
has zero determinant for $\alpha = \gamma = 0$. This requirement yields 
the following condition:
\begin{eqnarray}
C(\beta) - p\,q\,\tilde{C} (\beta) = \delta \sqrt{(1 - p^2)(1 - q^2)}.
\label{spinodal}
\end{eqnarray}
This line may be interpreted as a spinodal line for the reconstruction
transition. Its stable part describes a smooth boundary between reconstructed 
and unreconstructed regions. As we will see later part of this line is actually 
thermodynamically unstable.

At this point we should also discuss the symmetry properties of the surface and 
its free energy. Obviously the latter should be invariant under the 
transformations $p \to -p$, $q \to -q$ and $p,q \to q,p$. Our expression for 
the unreconstructed free energy satisfies all these requirements. The one for 
the reconstructed free energy is invariant under the last transformation and 
under $p,q \to -p,-q$, but it is not invariant under reflection of $p$ or $q$ 
alone.
A little thinking reveals that we have broken this symmetry by our 
identification of positive $p$ and $q$ with given arrow directions; if one 
identifies the opposite horizontal arrow direction with positive $p$ {\em and}
changes the sign of the actual horizontal polarization obviously our expression
for the free energy does not change. So to restore the required symmetry one
has to adopt the convention that positive $p$ and $q$ should correspond to the 
direction of the majority of the horizontal respectively vertical arrows. This 
does not look unreasonable: our mean field mean assumptions obviously work 
better the smaller the number of overturned arrows with respect to the fully 
polarized state. On the other hand, in stable reconstructed states the fraction 
of overturned arrows often is close to one half, so the mean field description 
used here may be less accurate than one might like. 

\section{Equilibrium Crystal Shapes}
\label{sec:ecs}

\subsection{Maxwell construction}

Unstable or metastable regions on a free energy surface can be stabilized by
applying a {\em Maxwell construction}. This amounts to connecting all pairs
of points on the free energy surface by tie lines, constructing the lower
envelope of the free energy surface combined with the set of all the tie lines 
and putting this in place of the original free energy surface, and repeating 
this procedure until it converges. The new points that are generated this way 
represent ``coexisting states" of two or three stable orientations.
In other words, the most favorable way for the crystal to realize surfaces of 
the corresponding orientations is by {\em faceting}, that is replacing these 
surfaces by a combination of two or three different surface elements with the 
same average orientation. Surface orientations that give rise to faceting do 
not appear in the equilibrium shape of the crystal; they are not 
thermodynamically stable.

\subsection{Orientations $0 \leq p=q \leq 1$}

The dashed lines in Fig. \ref{FIG05} show an example of the Maxwell 
construction applied to the orientations $0 \leq p=q \leq 1$.

{From} the stable free energy $f$ one finds the equilibrium shape
using \cite{henkim}:
\begin{eqnarray}
x=\frac{\partial f}{\partial p},
\,\,\,\,\,
y=\frac{\partial f}{\partial q}
\,\,\,\,\,\, {\rm and} \,\,\,\,\,\,
z=f-p \frac{\partial f }{\partial p} -q \frac{\partial f}{\partial q}.
\label{shape}
\end{eqnarray}
up to an arbitrary prefactor. These equations can be used to express $z$ 
as a function of $x$ and $y$. For the unreconstructed rounded parts of the 
surface one has to substitute Eq.\ (\ref{freeunr}) for $f$. 
This yields a shape equation of the form
\begin{equation}
z = \delta - C \sqrt{
\left( 1 +  \frac{x^2}{C^2} \right)
\left( 1 +  \frac{y^2}{C^2} \right)
}
\label{shapexy}
\end{equation}
where 
$C$ was defined in Eq.~(\ref{C}). 
{From} the equations (\ref{shape}) applied to the unreconstructed free 
energy one also finds that $x y = C^2 p q$; it then follows that the 
boundary of the $p = q = 1$ facet is simply given by a hyperbola:
\begin{eqnarray}
x y = C^2. 
\label{hyperbola}
\end{eqnarray}

Applying the Maxwell construction is not very practical since the surface 
free energy depends on two slope parameters. 
Instead one may obtain the equilibrium shape of the crystal by just applying
Eqs.\ (\ref{shape}) and discarding unstable wings, as described in 
\cite{henkim}.
E.g.\ all points with $z > 0$ are unstable, as they would require some 
non-convex surface part to reach them from the top facet, which is located at
$z=0$. Unstable wings can also occur in the region $z \leq 0$, as will be
shown in the numerical analysis of the equilibrium shape for generic $p$
and $q$.

Along the symmetry line $p=q$ one can use the analytic expressions 
Eqs. (\ref{recq=p}) and (\ref{unrecq=p}) for the reconstructed and 
unreconstructed free energies.
One finds that at low temperatures Eq.\ (\ref{shape}) gives  $z > 0$, 
except for the two facets which are then connected directly under a 
sharp edge as at $T=0$.
The shape along the symmetry line is then given by:
\begin{equation}
z = \left \{
\begin{array}{ccc}
0 & \,\,\,\,{\rm for}\,\,\,\, & |x| = |y| \leq \delta/2 \nonumber\\
\delta - 2 |x|  & \,\,\,\, {\rm for} \,\,\,\,& |x| = |y| \geq \delta/2 
\end{array}
\right.
\label{tltter}
\end{equation}

At higher temperatures in the vicinity of the $q = p = 1 $ facet some 
unreconstructed orientations become stable. The condition for this to 
happen is that the Maxwell construction connects $q = p = 0$ with 
$q = p = {\hat q} < 1$ (See Fig.\ \ref{FIG05} (b) for an example) or, 
equivalently, that in Eq. (\ref{shapexy}) a range of orientations will 
give rise to $z < 0$. Applying Eqs.\ (\ref{shape}) and (\ref{freeunr})  
for $p,q \to 1$ one finds that $z= \delta -2C(\beta)$. 
Equating this to zero one obtains the so-called 
edge rounding temperature $T_{\rm ER}$ from
\begin{equation}
\delta =
2 C(\beta_{\rm ER})
\label{Ter}
\end{equation}
For $T > T_{\rm ER}$ there is a range of stable unreconstructed orientations 
connected to the $(001)$ facet under a sharp edge, where the state $q = p =0$ 
coexists with a state $q = p = {\hat q}$. The value of ${\hat q}$ can be 
found either by applying the Maxwell construction or again from the 
requirement $z(x=y)=0$ in Eq. (\ref{shapexy}).
One finds:
\begin{eqnarray}
{\hat q} = \sqrt{\frac{\delta - C}{C}}.
\label{defq1}
\end{eqnarray}
For $T_{\rm ER}<T < T_{\rm s}$, with $T_{\rm s}$, given by Eq. (\ref{tsmooth}), 
the shape profile is given by:
\begin{equation}
z = \left \{
\begin{array}{ccc}
0 &\,\,\,\, {\rm for} \,\,\,\,& |x| = |y| \leq {\hat x} \nonumber\\
\delta - C - x^2 /C  &\,\,\,\, {\rm for}\,\,\,\, & {\hat x} \leq |x| 
= |y| \leq C \nonumber\\
\delta - 2 |x|  &\,\,\,\, {\rm for}\,\,\,\, & |x| = |y| \geq C
\end{array}
\right.
\label{tgtter}
\end{equation}
with ${\hat x} = C {\hat q}$. 
Notice that $z(x)$ has a jump in the first derivative at $x = {\hat x}$
(sharp edge) and in the second derivative for $x = C$ (smooth edge).

As pointed out in the previous section, above the temperature $T_{\rm s}$ 
(see Eq. (\ref{tsmooth})) there is a range of temperatures where
the reconstructed free energy (\ref{recq=p}) becomes stable.
Applying the Eqs.\ (\ref{shape}) to the reconstructed free energy one finds
the following shape profile:
\begin{equation}
z = \left \{
\begin{array}{ccc}
0 & {\rm for} & |x| = |y| \leq x_f \nonumber\\
A (x_f - x)^2 & {\rm for} & x_f \leq |x| = |y| \leq x_0 
\nonumber \\
\delta -C - x^2 /C & {\rm for} & x_0 \leq |x| = |y| \leq C
\nonumber \\
\delta - 2 |x|  &\,\,\,\, {\rm for}\,\,\,\, & |x| = |y| \geq C
\\
\end{array}
\right.
\label{tgtts}
\end{equation}
with: 
\begin{eqnarray}
x_f = \sqrt{(\delta - C)(\delta - \tilde{C})},
\end{eqnarray}
\begin{eqnarray}
x_0 = C \sqrt{\frac{\delta - C}{\delta - \tilde{C}}},
\end{eqnarray}
\begin{eqnarray}
A = - \frac{1}{C + \tilde{C} - \delta}.
\end{eqnarray}
The rounded regions in the range $x_f \leq |x| = |y| \leq x_0$ are thus
reconstructed and connected smoothly to the top facet (which is also 
reconstructed) and to the unreconstructed rounded regions.
As remarked previously, by approaching the boundary between reconstructed 
and unreconstructed regions from the reconstructed side i.e. for $x = y 
\to x_0^-$, one finds that the reconstruction order parameter $\alpha$ 
vanishes continuously with a mean-field exponent $1/2$, as is easily seen 
from equation (\ref{alq=p}) combined with $x = C q$. 
At this point also the shape profile is singular, as the second derivative
of $z(x)$ has a jump at $x=x_0$. 
Notice that the shape is parabolic in the vicinity of the $(001)$ and $(011)$
facets, i.e. $z - z' \sim (x - x')^2$. This is typically a mean-field result,
the actual exponent ought to be $3/2$ instead of $2$, as it was found in 
exactly solved models \cite{henkim}.
Finally, at the temperature $T_{\rm MF}$ 
(see (\ref{Tmf})) one has $x_f$, $x_0 \to 0$ confirming the simultaneous 
occurrence of a roughening and  a deconstruction transition  discussed 
already in the previous section. For $T \geq T_{\rm MF}$ the rounded part 
of the crystal is described entirely by (\ref{shapexy}).

\begin{figure}[ht]
\centerline{
\psfig{file=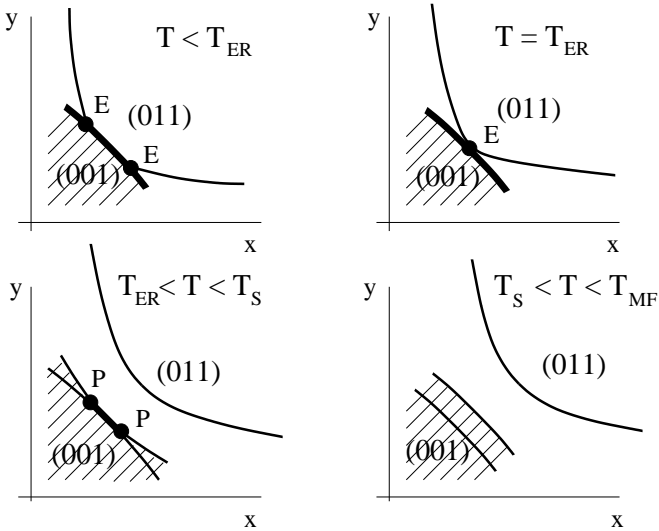,height=7cm}}
\vskip 0.2truecm
\caption{
Thermal evolution of the crystal shape (top view) in the vicinity of 
the line $x = y$. Thick lines denote sharp edges in the shape. For a 
detailed description see text. Above the temperature $T_{\rm MF}$ the 
$(001)$ facet, as well as the reconstructed orientations disappear 
from the crystal, and the global shape is described by Eq. (\ref{shapexy}).
}
\label{FIG06}
\end{figure}

Figure \ref{FIG06} shows schematically the top view of the equilibrium
crystal shape in the vicinity of the line $x = y$ at different 
temperatures. At low temperatures $T < T_{\rm ER}$ the $(001)$ and
$(011)$ facets are connected to each other along the segment $EE$.
The $(011)$ facet boundary, starting from the points $E$ is described 
by the two branches of the hyperbola given in Eq. (\ref{hyperbola}).
The coordinates of the points E can be found exactly from the 
intersection of (\ref{hyperbola}) with the constant height contour 
(\ref{shapexy}) for $z=0$, where the hyperbola intersects the $(001)$
facet. One finds:
\begin{eqnarray}
x_E = \sqrt{\frac{\delta^2}{2} - C^2 + 
\delta \sqrt{\frac{\delta^2}{4} - C^2}}
\label{xE} \\
y_E = \sqrt{\frac{\delta^2}{2} - C^2 - 
\delta \sqrt{\frac{\delta^2}{4} - C^2}}
\label{yE}
\end{eqnarray}
The two points of coordinates $(x_E, y_E)$ and $(y_E, x_E)$ merge into a
single one at the edge rounding temperature ($T = T_{\rm ER}$) where
$x_E = y_E = \delta/2$. For $T < T_{\rm ER}$ the boundary between the two
points $E$ is described by the equation $x + y = \delta$, as for $T=0$.
For $T_{\rm ER} < T < T_s$ the $(001)$ and $(011)$ facets are completely
separated from each other. The full analysis for generic $p$ and $q$ shows
(more details later) that the $(001)$ facet edge is sharp only between
the two points $P$ shown in Fig. \ref{FIG06}(c). Beyond these
points one finds rounded reconstructed regions and a smooth $(001)$-facet
edge. Finally for $T_s < T < T_{\rm MF}$ the $(001)$ facet edge is 
surrounded by curved reconstructed orientations and it connects smoothly to 
the rounded regions along its full circumference in the whole crystal.

\subsection{General shape}

For generic $p$ and $q$ we have calculated the mean field equilibrium shape 
numerically, except outside the boundary given by Eq. (\ref{spinodal}), where 
the unreconstructed state prevails and the equilibrium shape is given by 
Eq. (\ref{shapexy}).

\begin{figure}[ht]
\centerline{
\psfig{file=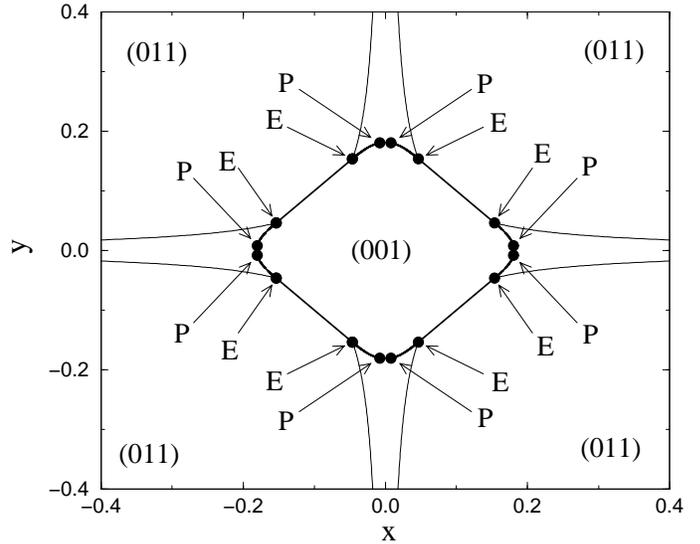,height=7.5cm}}
\vskip 0.2truecm
\caption{Equilibrium crystal shape (top view) for $\epsilon = -2$, 
$\delta = 0.2$ and for $T = 1.0$, i.e. below the edge rounding temperature 
($T_{\rm ER} \approx 1.065$). Thick and thin lines denote sharp and smooth 
facet edges respectively. The boundaries of the $(011)$ facets with the rounded 
regions are always smooth. For a detailed description of the points E and P
see the text.
}
\label{FIG07}
\end{figure}

Figure \ref{FIG07} shows a top view of the equilibrium crystal shape
for $\epsilon = -2$, $\delta = 0.2$ and for a temperature $T = 1.0$, 
i.e.  below the edge rounding temperature $T_{ER} \approx 1.065$. 
Thick lines mark those parts of the $(001)$ facet boundary which are 
sharp. The EE segments are sharp and straight as at $T = 0$. 
The parts PE are also sharp, here the $(001)$ facet is connected to
unreconstructed rounded regions. The parts PP are smooth and the $(001)$ 
facet is connected to reconstructed rounded regions. The shape of the 
crystal around the points P, in the vicinity of the axes $x=0$ and $y=0$,
discussed later. The boundaries of the $(011)$ facets with rounded regions 
are always smooth and marked by thin lines originating from the points E.

\begin{figure}[ht]
\centerline{
\psfig{file=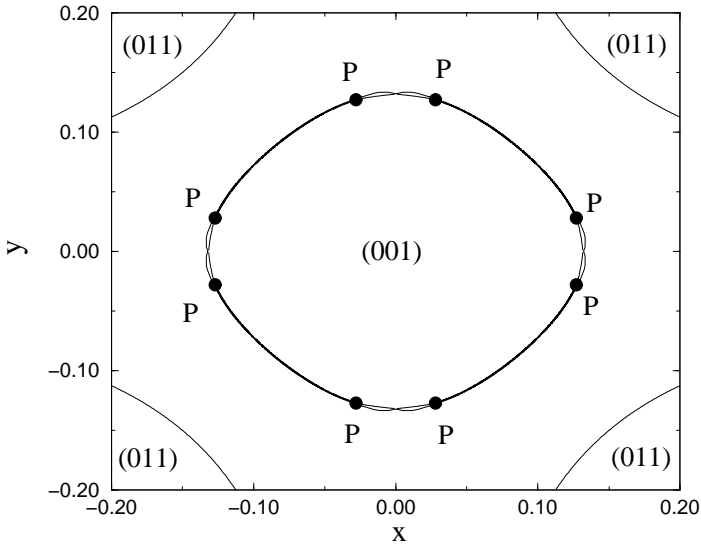,height=7.5cm}}
\vskip 0.2truecm
\caption{As in Fig. \ref{FIG07} for a temperature $T = 1.25$, i.e. above the 
edge rounding temperature. The part of the facet edge between the two points 
P, indicated with a thick solid line, is sharp.
Between two points P and close to the axes $x=0$, $y=0$ there are two wings of 
reconstructed rounded regions, which are shown in some detail for 
$T=1.111\dots$ in Fig. \ref{FIG09}.
}
\label{FIG08}
\end{figure}

\begin{figure}[ht]
\centerline{
\psfig{file=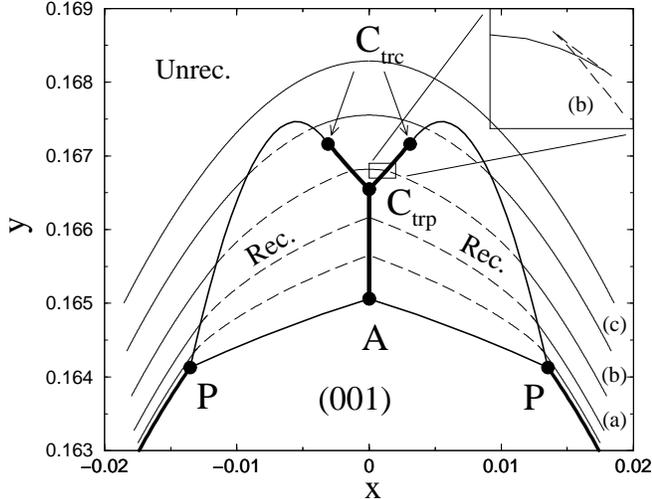,height=7cm}}
\vskip 0.2truecm
\caption{Top view of an enlargement of the ECS close to the intersection
of the edge of the $(001)$ facet with the axis $x=0$. For $\epsilon = -2$, 
$\delta = 0.2$ and $T=1.111\ldots$. The thinnest lines denote contours for 
fixed values of $z$. The $(001)$ facet is located at $z=0$ and the other 
lines correspond to $z=-0.0001$, $z=-0.0003$ (a), $z=-0.0007$ (b), 
$z=-0.00125$ (c) and $z=-0.0018$.
}
\label{FIG09}
\end{figure}

Figure \ref{FIG08} shows the ECS for $T = 1.25$, i.e. above the 
edge rounding temperature. Now all facets are separated from each other 
by rounded regions. The (001) facet boundary is sharp in the part marked
by thick solid lines; here the (001) facet is connected to unreconstructed 
rounded regions. Figure \ref{FIG09} shows a blow-up of the equilibrium 
crystal shape in the vicinity of the $x = 0$ axis for $T=1.111\ldots$.
The area below the P-A-P curve belongs to the $(001)$ facet. 
As in Figs.\ \ref{FIG07} and \ref{FIG08} thick solid lines denote sharp 
edges. Thinner lines are smooth boundaries, either facet edges (as the 
segments A-P) or boundaries between the reconstructed and unreconstructed 
regions (curve P-C$_{\rm trc}$).
To facilitate the description of the crystal shape we have plotted some 
contour lines corresponding to constant values of $z$. They have been drawn 
solid in the unreconstructed part of the crystal and dashed in the 
reconstructed regions. The $(001)$ facet has $z=0$ and the contours correspond 
to the respective values $z=-0.0001$, $z=-0.0003$, $z=-0.0007$, $z=-0.00125$
and $z=-0.0018$.
Starting from the contour line (a) at $x \approx 0.15$, i.e. in the 
unreconstructed rounded region and shifting towards smaller values of $x$ 
one encounters first the P-C$_{\rm trc}$ boundary which separates the 
unreconstructed from the reconstructed region; the crystal shape at this 
boundary is smooth, in the sense that the first derivatives of $z(x,y)$
as function of $x$ and $y$ are continuous, but higher derivatives are not.
Proceeding further along the contour line (a) one terminates in the 
A-C$_{\rm trp}$ segment, which is a sharp ridge in the rounded part of
the crystal \cite{note2}.
For the contour line (b), which corresponds to a lower value of $z$, a 
similar behavior as for (a) is found, except that the reconstructed 
region terminates in the segment C$_{\rm trp}$-C$_{\rm trc}$ and not at
$x = 0$. 
The inset of Fig. \ref{FIG09} shows an enlargement of the contour line
(b) in the area around C$_{\rm trp}$-C$_{\rm trc}$. The contour line
is obtained from the Legendre transform of the full free energy, without
applying the Maxwell construction. 
This procedure produces an unstable wing which must be eliminated from the 
shape. Unstable wings are rather easy to be identified and indicate the
existence of a sharp boundary in the crystal shape. The contour (c) crosses
the boundary in two points, and in both points the boundary is smooth;
in this case we did not observe any unstable wing along the line.
The point C$_{\rm trc}$ marks the end of the first order line and it is
a {\it tricritical} point. The point C$_{\rm trp}$ where the three first 
order lines merge is a {\it triple} point. Notice that the reconstructed 
region is rather small; this area is hardly visible on the scales of
Figs. \ref{FIG07} and \ref{FIG08}. As the temperature is increased the
reconstructed areas occupy a larger portion of the crystal surface.

The coordinates of the point P can be found analytically from the 
intersection of Eq. (\ref{spinodal}) with the unreconstructed crystal 
shape Eq. (\ref{shapexy}) with $z = 0$.
One obtains:
\end{multicols}\widetext

\begin{equation}
x_P = \sqrt{
\frac{\delta^2 - C^2 - {\tilde C}^2  
- \sqrt{
\left[ \delta^2 - (C + {\tilde C})^2 \right] 
\left[ \delta^2 - (C - {\tilde C})^2 \right]
}
} 
2}
\end{equation}
\begin{equation}
y_P = \sqrt{
\frac{\delta^2 - C^2 - {\tilde C}^2 
+ \sqrt{
\left[ \delta^2 - (C + {\tilde C})^2 \right] 
\left[ \delta^2 - (C - {\tilde C})^2 \right]
}
} 
2}
\end{equation}
\begin{multicols}{2}\narrowtext

At a temperature $T_s$ obtained from the condition (\ref{tsmooth})
the two points P merge into a single point ($x_P = y_P$). Above 
$T_s$ the points P disappear and the whole edge of the $(001)$ facet 
connects smoothly to the rounded regions surrounding the (001) facet.
Still in the vicinity of the axes $x = y = 0$ one has a tricritical 
and triple points as in Fig. \ref{FIG09}. As discussed before there 
is a temperature $T_{\rm MF}$ where the $(001)$ facet vanishes 
(roughening) and simultaneously the reconstruction disappears.

\begin{figure}[ht]
\centerline{
\psfig{file=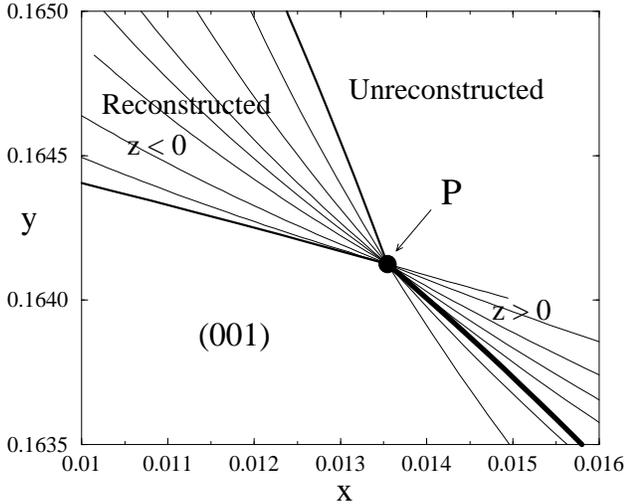,height=7cm}}
\vskip 0.2truecm
\caption{
Enlargement of the top view of the ECS around the point P ($T = 1.111\ldots$). 
Thin lines are trajectories obtained at constant ratios $p/q$. The fact that
all these intersect in P indicates that P is a {\it conical} point.
}
\label{FIG10}
\end{figure}

We conclude the discussion of the crystal shape with an analysis of the
properties of the point P, which turn out to be very interesting. 
A numerical solution of Eqs.\ (\ref{mfree}) and (\ref{shape}) leads to the 
results illustrated in Figure \ref{FIG10}. Several lines connecting points 
with equal values of $p/q$  intersect each other in the common point $P$, 
with a wide range of values for $p$ and $q$ assumed in this point. This 
corresponds to the geometry of the tip of a cone, where one also has tangent
planes of a wide range of orientations. Therefore we will refer to P as
as a {\em conical point}.
An analogous point was found in a non-staggered version of the six
vertex model \cite{joeldj}, which is exactly solvable. Figure \ref{FIG10} 
shows an enlargement of the region around the point P. 
The temperature is the same as in Fig. \ref{FIG09}. The 
thick solid line denotes the sharp (first order) boundary which opens up in 
P into two continuous boundaries. The lower one of these is the smooth $(001)$ 
facet edge and the upper one is the boundary between reconstructed and 
unreconstructed rounded regions. 
The thin lines are trajectories obtained by keeping the ratio $q/p$ 
constant, which indeed are seen to all intersect each other in P. 
Notice that P is at the facet edge, therefore has $z=0$. The parts
of the lines to the left of P correspond to $z < 0$ and are stable. 
The parts to the right of P have positive $z$ and describe an unstable wing,
which has to be eliminated from the equilibrium shape.
In some sense the point P is highly degenerate since a continuous set 
of values of $q$ and $p$ coexist. 
In a plot of the free energy vs. orientation, i.e. of $f$ as a function of 
$p$ and $q$, this is reflected in the form of a flat area for a certain 
range of orientations, namely those physically realizable orientations that 
are tangent at the conical point and all their convex combinations. The 
existence of such a flat area is a direct consequence of Eq.(\ref{shape}); for 
different points on the $f(p,q)$ surface to give rise to the same $x$, $y$ 
and $z$ they have to lie in a common plane. From a physical point of view the 
existence of such a flat area in the $f(p,q)$ surface looks extremely 
surprising. It looks like having the possibility for a finite range of step 
densities for which the step dependent part of the free energy is strictly 
proportional to the step density, but gets no contributions whatever from 
interactions between the steps.
It should be stressed however that so far almost all our evidence for the 
conical character of P is numerical. Although the accuracy of our calculations 
is very high and all our results indicate conic behavior, we cannot exclude 
the possibility that on an even finer scale the lines of various $p/q$ will 
not intersect exactly and in fact the single conical point P will extend into 
some folds and creases, the extension of which certainly has to be very small.
However, there is one important piece of analytic support for a strictly 
conic behavior of P, which is the behavior of $f(p,q)$ at the temperature 
$T_s$. Here the two points P as drawn in Fig.\ \ref{FIG07} 
merge into a single point.
According to Eq. (\ref{recq=p}) all orientations $p=q$ with 
$0< q <q_0=\sqrt{\tilde{C}(\beta_s)/C(\beta_s)}$ indeed join in the 
point P. This is a limiting case of a conical point, where the opening angle 
of the cone becomes zero. There seems as little physical reason for the 
coincidence of all these orientations in P as in the general case, but of 
course this does not provide a proof for conical behavior in general.
 
Notice that in a previous mean-field calculation \cite{rapid} the two
facets $(101)$ and $(011)$ where found connected under a sharp edge
at large $p$ and $q$ values. This is an artefact of that approach, which has 
the following drawbacks:
In that description the starting point of the mean-field calculation
was the Manhattan lattice. 
Therefore the description of the orientations near the $(011)$ type facets
is less accurate than in the present method. But it now turns out that,
besides the facet orientations, these are the only ones that are actually
found in the equilibrium shapes at low temperatures. In addition both mean
field approximations give the same result for the free energy of the
(001)-facet, so no accuracy is lost there.
Finally the present mean field approximation is much more amenable to analytic 
treatment and for instance symmetries under sublattice exchange and/or arrow 
reversal remain much more transparent.

\section{Discussion}
\label{sec:disc}

The model studied in this paper is expected to give a fairly good description
of surfaces of ionic crystals of bcc type, such as CsCl. We find an equilibrium 
shape with extended facets and mostly sharp edges for the (001) facet at low
temperatures. The shape is very rich and exhibits critical points of
various nature, such as triple, tricritical and conical points. This richness 
stems from the fact that, besides the facet edges of the (001) and (011) 
facets the model also has boundaries between reconstructed and unreconstructed 
regions. The interplay between the former and the latter produces such a rich 
phase diagram. There exists a fair number of other models for equilibrium 
shapes which have been solved either in mean-field approximation or by 
transfer matrix methods, but we are not aware of cases where the shape is 
as complex as in the case presented here. Particularly striking is the 
likely appearance of conical points, which, as pointed out above, have a 
special degeneracy, i.e. a continuous set of different slopes $p$
and $q$ coexist in such points. Indicating the coordinates of the conical 
point with ($x_P$, $y_P$, $z_P$) we have from Eq.\ (\ref{shape}): 
$f(p,q) = z_P + p x_P + q y_P$, i.e. the coexisting set of orientations 
$p$ and $q$ is such that their surface free energy has the form of a planar 
facet in the $f(p,q,)$ surface.
Notice that the opposite situation is much more common and easier to understan
d physically, i.e. a singular point of conical type in $f(p,q)$ produces a 
facet in $z(x,y)$. 

Finally, it is interesting to point out that shapes of $NaCl$ crystals 
in thermal equilibrium were investigated by Heyraud and M\'etois \cite{salt}. 
They found that the shape of the crystal is a perfect cube up to a 
temperature $T \approx 620^\circ$ and that the roughening of the crystal
starts from the corners of the cube and extends towards the edges. 
Hence, it seems that the presence of sharp facet boundaries, persisting at 
high temperature, is a common feature of crystals of ionic type \cite{wortis1}.

{\bf Acknowledgements} - 
It is a pleasure to thank D.\ J.\ Bukman, J.\ W.\ M.\ Frenken, G.\ Mazzeo,
J.\ D.\ Shore and M.\ Wortis for valuable comments and
discussions. E.\ C.\ would like to thank M.\ Wortis for his kind 
hospitality at the Physics Department of the Simon Fraser University, 
where part of this work was done.

\section*{Appendix I: The free energy of a step on the $(011)$ facet}

Excitations on the $(011)$ facet of the  six vertex models are easy 
to study. 
For this orientation the vertex lattice is fully polarized with all arrows 
pointing say, up and to the right, as shown in Fig.~\ref{FIG03}
No closed loops of reversed arrows are possible since 
all reversed arrows point down or to the left.
The only possible excitations are infinitely long steps (or steps 
between two boundaries). The steps
can follow only two directions and cannot intersect themselves so that	
the step free energy can be calculated exactly.

In order to calculate the free energy of an isolated step we introduce
the $2 \times 2$ matrix $G(k)$ as the ``lattice Green function"
in momentum space for straight step segments.
The matrix is written in the following form:
\begin{eqnarray}
G(k) =       
\left(
\begin{array}{cc}
G_{\times \times}(k) \,\,\,& \,\,\, G_{\times \bullet}(k) \\
G_{\bullet \times}(k) \,\,\,& \,\,\, G_{\bullet \bullet}(k)
\end{array}
\right) 
\label{defg}
\end{eqnarray}
where $\times$ and $\bullet$ denote the two different types of lattice 
points, defining the two sublattices of the staggered six vertex model, 
as indicated in Fig \ref{FIG03}. In this notation $G_{\times\times} (k)$, for 
instance, stands for the sum of the Boltzmann weights of straight horizontal 
segments between two points of type $\times$ of length $n$, multiplied by 
$e^{ikn}$. The Boltzmann weight counts the energies of the changed vertices 
between the end points of the segment and, by convention, the energy of the 
turn at the beginning of the segment. 
The straight segments of steps have Boltzmann weights $e^{-2 \beta \delta}$ 
and $ e^{2\beta \delta}$ in lattice points of type $\bullet$ and $\times$, 
respectively. Turns have weights $e^{\beta \epsilon}$ in $\bullet$ and
$e^{\beta (\epsilon -2 \delta)}$ in $\times$. For $G_{\times \times}(k)$ this 
yields
\begin{eqnarray}
G_{\times \times}(k) = e^{\beta (\epsilon + 2 \delta)} e^{-2\beta\delta }
\sum_{n=1}^{\infty} e^{2 i k n} = \frac{e^{\beta \epsilon }}{e^{-2 i k}-1}
\end{eqnarray}
The matrix (\ref{defg}) is given by
\begin{eqnarray}
G(k) = \frac{e^{\beta \epsilon }}{e^{-2 i k}-1}
\left(
\begin{array}{cc}
1 & e^{-i k} \\
e^{-i k - 2\beta \delta} & e^{
2 \beta \delta}.
\end{array}
\right) 
\end{eqnarray}

The full lattice Green function is obtained by considering all possible 
alternating sequences of horizontal and vertical segments; the
contributions of the latter can be accounted for by the same matrix 
$G$ with an argument $k_v$ instead of $k_h$, for the horizontal
steps. One obtains
\begin{eqnarray}
{\widehat G} ( \vec k ) &=& G(k_h) + G(k_v) + G(k_h) G(k_v) + 
G(k_v) G(k_h) + \ldots
\nonumber \\
&=& \left(1 +G(k_v)\right) \frac 1
{ 1 - G(k_h) G(k_v) }
\left(1+G(k_ h)\right) - 1 
\end{eqnarray}
where $h$ and $v$ denote the two orthogonal directions of the axes of 
the vertex lattice.

For the calculation of the partition function as function of step orientation 
and step length one transforms back to real space:
\begin{eqnarray}
Z (L,\phi) = \sum_{ij} \int_{-\pi}^{+\pi} \, \frac{d \vec k}{4 \pi^2} \, 
e^{-i k_h L_h} \, e^{-i k_v L_v} \, {\widehat G}_{ij}(\vec k)
\label{integral}
\end{eqnarray}
Here $L$ denotes the total length and $L_h = L \cos \phi$, 
$L_v = L \sin \phi$.
Notice that we sum over all possible entries of ${\widehat G}(\vec k)$, 
which is a sum over all possible types of initial and final positions 
($i,j=\{{\times ,\bullet}\}$).
Setting $z = e^{i k_h}$, one can perform the integration over $k_h$ exactly. It
becomes an integration along the unit circle in the complex $z$ plane. 
For the remaining integral over $k_v$ one can use the saddle point 
approximation, and in the limit $L \to \infty$ one obtains an expression of 
the type:
\begin{eqnarray}
Z(L,\phi) \sim e^{-L \beta f^{(011)}_s(\phi)}
\label{partfun}
\end{eqnarray}
with $f^{(011)}_s(\phi)$ the step free energy per unit of length. For a step 
running under an angle $\phi =\pi/4$ with the principal axes of the vertex 
lattice the calculation is particularly simple; the step free energy is:
\begin{eqnarray}
\beta f_s^{(011)}\left(\frac{\pi}{4}\right) = - \frac{1}{\sqrt 2} \log
\left[ \left(1 + e^{\beta \epsilon} \right) \left(1+ e^{\beta (\epsilon +
2 \delta)} \right) \right]
\label{step011}
\end{eqnarray}
The cause that this quantity is negative is that, for convenience, we have 
calculated the increase of the free energy due to the presence of a step on 
the $(011)$ facet, {\em per unit of projected area onto the $(001)$ facet}.
Calculating the same quantity per unit of projected area onto the $(011)$ 
orientation one finds a positive step free energy.
For a step of arbitrary orientation $\phi$ the calculation becomes somewhat 
more complicated. The resulting step free energy is of the form
\begin{eqnarray}
\beta f_s^{(011)}(\phi) = - \sqrt{\cos \phi \sin \phi} \,\, 
\left(1+ e^{2 \beta \delta} \right) \,\, e^{\beta \epsilon}
\label{step011gen}
\end{eqnarray}

{From} the single step free energy one obtains the surface free energy in the 
neighborhood of the $(011)$ facet taking into account only non-interacting 
steps, which gives a contribution linear in the step density \cite{near1}. 
One finds:
\begin{eqnarray}
f(p,q) \approx \delta - \frac{\sqrt{(1 -|q|)(1 - |p|)}}{2 \beta}
\left(1+ e^{2 \beta \delta} \right) \,\, e^{\beta \epsilon}
\label{exactfr011}
\end{eqnarray}
which is valid to lowest order in an expansion around the 
$(011)$ facet where $q=p=1$.
Comparing (\ref{exactfr011}) with (\ref{freeunr}) one concludes that the 
mean field free energy reproduces the exact surface free energy in the 
neighborhood of the $(011)$ facet 
to linear order in the step density.

\section*{Appendix II: The limit $\delta \to 0$}

In the limit $\delta \to 0$ the six vertex model has been solved exactly
\cite{baxter} and the exact solution can be compared with the mean field 
results in the limit of small $e^{\beta \epsilon}$.
For small values of $q$ and $p$, the exact solution yields
\cite{irmgard}:
\begin{equation}
f(p,q) = f(p=0,q=0) + 
\frac{r}{4 \beta}
\left[ p^2 + q^2  \right]
\label{exac1}
\end{equation}
with $\cos r \equiv  1 - e^{2 \beta \epsilon}/2$.
For small $e^{\beta \epsilon}$ one has (see Ref. \cite{baxter}, 
Eq. (8.11.7)):
\begin{eqnarray}
f(p=0,q=0) = - \frac{e^{\beta \epsilon}}{2 \beta} + O\left(
e^{3 \beta \epsilon/2} \right)
\label{exac2}
\end{eqnarray}
Combining (\ref{exac1}) with (\ref{exac2}), and noticing that at lowest order
in $e^{\beta \epsilon}$ one has $r = e^{\beta \epsilon}$, one obtains as 
expression for the exact free energy
\begin{eqnarray}
f(p,q) = -\frac{e^{\beta \epsilon}}{2 \beta} \left(
1 - \frac{p^2 + q^2}{2},
\right)
\end{eqnarray}
which agrees with the mean field free energy (\ref{freeunr}) for $\delta = 0$
and to lowest order in $p$ and $q$.

\end{multicols}


\begin{references}
\bibitem{notea} Notice that in the case of faceting a given
unstable or metastable surface orientation may decay in general
into a combination of three surface orientations.
\bibitem{lastnote} Crystallographically NaCl has an fcc lattice;
however, if one ignores the identities of the different ions one
obtains a simple cubic lattice.
\bibitem{near1} If one keeps either $|p|$ or $|q|$ strictly equal to unity
and moves the other variable slightly away from it, $f(p,q)$ does not
change. This is understood easily, because in that case the steps are
perfectly straight and do not change the free energy, as argued in 
Section \ref{subsec:gsp}.
\end{references}
\end{document}